\documentclass[preprint,12pt,a4paper]{elsarticle}
\usepackage{natbib} % Required for Elsevier-style bibliography
\usepackage{amssymb}
\usepackage{amsmath} % Include the amsmath package
\usepackage{caption}
\usepackage{graphicx}
\usepackage{xcolor}
\usepackage[toc,page]{appendix}

\usepackage[colorlinks,allcolors=blue]{hyperref}
\usepackage{subcaption}
\usepackage{adjustbox}
\usepackage{svg}

\definecolor{bleudefrance}{rgb}{0.19, 0.55, 0.91}

\usepackage{color}

\journal{SoftwareX}

\usepackage{siunitx}

\begin{document}

\begin{frontmatter}



\title{Discrete Element Simulations of particles interacting via capillary forces using \textit{MercuryDPM}}
% \title{CapDEM: DEM simulations of Capillary force interactions in particle systems using \textit{MercuryDPM}}

\author{Meysam Bagheri$^*$, Sudeshna Roy, Thorsten P\"oschel}
\address{
Institute for Multiscale Simulation, Friedrich-Alexander-Universit\"at Erlangen-N\"urnberg, Erlangen, Germany}
\cortext[correspondingauthor]{Correspondence: meysam.bagheri@fau.de (M.B.)}

\begin{abstract}
We present the implementation of two advanced capillary bridge approximations within the Discrete Element Method (DEM) framework of the open-source code \textit{MercuryDPM}. While \textit{MercuryDPM} already includes a simplified version of the Willett approximation, our work involves implementing both the classical Willett approximation and the recently published Bagheri approximation in \textit{MercuryDPM}. Through detailed descriptions and illustrative simulations using a two-particle collision model, we demonstrate the enhanced accuracy and capabilities of these approximations in capturing the complex dynamics of wet granular matter.
\end{abstract}

\end{frontmatter}

\section*{Metadata}
\begin{table}[!h]
\begin{tabular}{|l|p{6.5cm}|p{6.5cm}|}
\hline
C1 & Current code version & MercuryDPM Master Version \\
\hline
C2 & Permanent link to code/repository used for this code version & \href{https://github.com/meysam-bagheri/MercuryDPM}{https://github.com/meysam-bagheri/MercuryDPM} \\
\hline
C3  & Permanent link to Reproducible Capsule & \\
\hline
C4 & Legal Code License   & BSD 3-clause \\
\hline
C5 & Code versioning system used & git \\
\hline
C6 & Software code languages, tools, and services used & C++ \\
\hline
C7 & Compilation requirements, operating environments \& dependencies & {Linux/Windows/macOS \newline see \href{https://www.mercurydpm.org/}{https://www.mercurydpm.org/}} \\
\hline
C8 & If available Link to developer documentation/manual & \href{https://github.com/meysam-bagheri/MercuryDPM/blob/master/README.md}{https://github.com/meysam-bagheri/MercuryDPM/blob/master
/README.md} \newline \href{https://www.mercurydpm.org/documentation}{https://www.mercurydpm.org/
documentation} \\
\hline
C9 & Support email for questions & \href{mailto:meysam.bagheri@fau.de}{meysam.bagheri@fau.de}; \href{mailto:sudeshna.roy@fau.de}{sudeshna.roy@fau.de}\\
\hline
\end{tabular}
\label{codeMetadata} 
\end{table}

% %% \linenumbers
% \section*{Metadata}
% \label{}
% \begin{table}[htbp]
% % \caption{DEM simulation parameters.}
% \label{tab:material_parameters}
% \begin{center}
% \begin{tabular}{p{0.4\textwidth} p{0.55\textwidth}} % Adjust column widths as needed
% \hline
% \textbf{Description} & \textbf{Details} \\ [2pt] % Add a header row if needed
% \hline
% {Current code version} & MercuryDPM Master Version \\
% Permanent link to code/repository used for this code version & 
% \href{https://github.com/meysam-bagheri/MercuryDPM}{https://github.com/meysam-bagheri/MercuryDPM}\\
% % {Code Ocean compute capsule} & \\
% {Legal Code License} & BSD 3-clause \\
% {Code versioning system used} & git\\
% {Software code languages, tools, and services used} & C++\\
% {Compilation requirements, operating environments \& dependencies} & {Linux/Windows/macOS \newline see \href{https://www.mercurydpm.org/}{https://www.mercurydpm.org/}}\\
% {Support email for questions} & \href{mailto:meysam.bagheri@fau.de}{meysam.bagheri@fau.de}; \href{mailto:sudeshna.roy@fau.de}{sudeshna.roy@fau.de}\\[2pt] 
% \hline                
% \end{tabular}
% \end{center}
% \end{table}

\section{Motivation and significance}
\begin{figure}
    \centering
    \includegraphics[width=10cm]{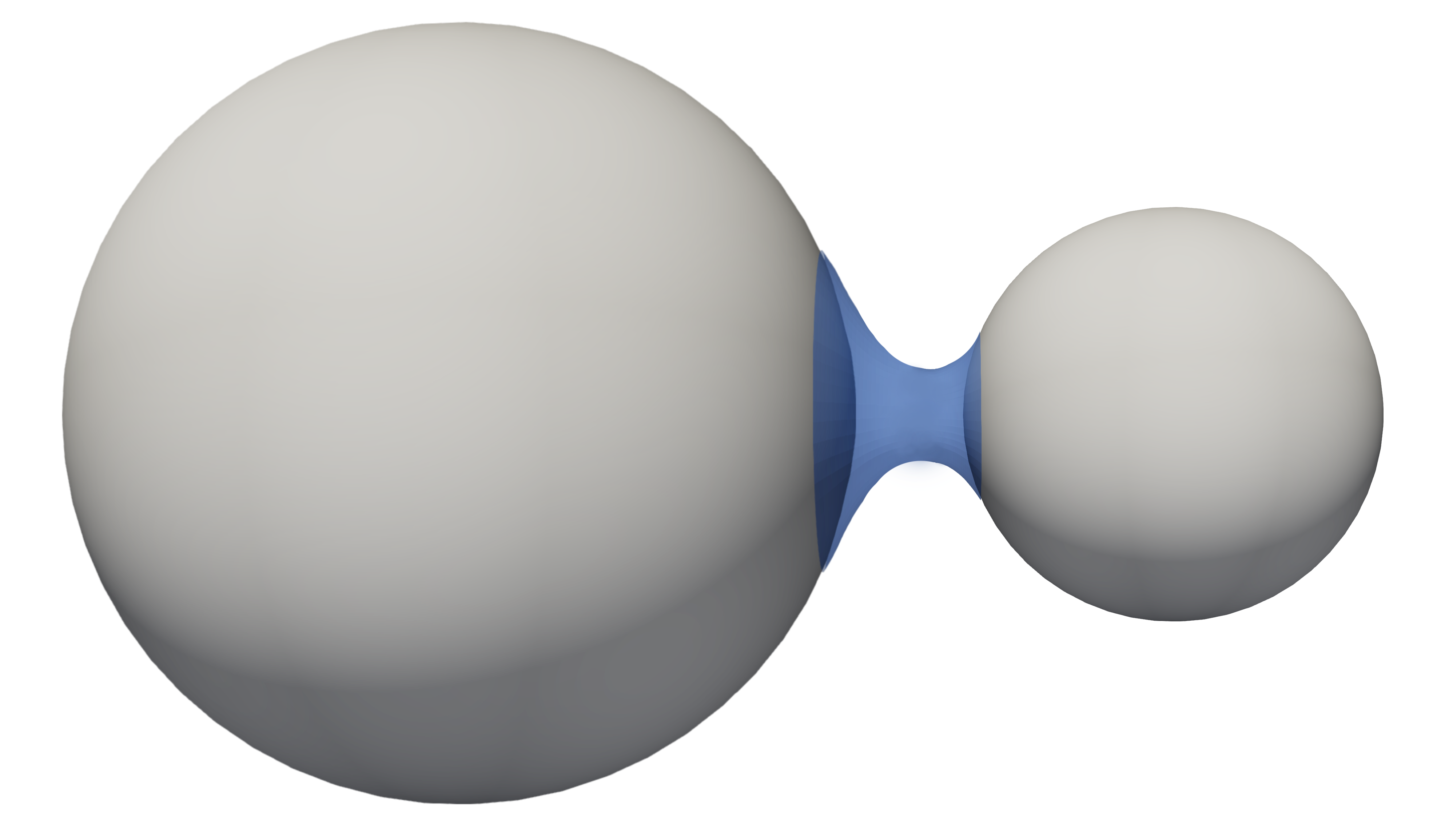}
    \caption{Capillary bridge spanning between two unequal-sized particles.}
    \label{fig:lb}
\end{figure}

In the study of particulate systems, understanding the interactions between particles in the presence of liquid phases is crucial for a variety of industrial and scientific applications \cite{scheel2008morphological,salimi2020undrained}. One of the key phenomena in such systems is the formation of liquid bridges between particles, which causes forces due to capillarity \cite{yang2021capillary}. These interactions can significantly influence the mechanical properties and behavior of granular materials \cite{radjai2009bond,roy2016micro,johanson2003relationship}. Liquid capillary bridges play a vital role in many processes, in the fields of soil mechanics and powder technology \cite{mckenna1989theoretical,price2002influence}. Accurately approximating these forces is therefore essential for reliable simulations of wet granular materials. Incorporating these complex interactions into DEM simulations poses significant challenges. The na\"{\i}ve way to account for capillary forces between particles linked with a liquid bridge (see \autoref{fig:lb}) would require solving the Young-Laplace equation for the capillary force in each time step of the DEM integration, which is certainly not feasible. Therefore, fit formulae were elaborated which approximate the capillary force as a function of the particle and liquid properties. The evaluation of these fit formulae is by orders of magnitude more efficient than the numerical solution of the Young-Laplace equation. The fit formulae by Willett et al. \cite{willett2000capillary} and Bagheri et al. \cite{bagheri2024approximate} have been shown to be highly accurate, where the latter delivers also the surface area of a liquid bridge which is needed when considering evaporation processes in wet granulate. By now, DEM packages rely on more simple and less accurate formulae. 

The current article reports the implementation of highly accurate descriptions of capillary forces in the DEM package \textit{MercuryDPM}. \textit{MercuryDPM} is a versatile open-source code designed for simulating the dynamics of particulate materials using DEM \cite{weinhart2020fast,thornton2023recent}. Its flexibility and extensibility make it an ideal tool for simulations of large scale granular systems
%implementing and testing new physical models, including those for liquid capillary bridges. 
By enhancing \textit{MercuryDPM} with accurate liquid bridge approximations, researchers and engineers can gain deeper insights into the behavior of wet granular systems and improve the design and optimization of related processes.

The Bagheri approximation, originally formulated for equal-sized spheres \cite{bagheri2024approximate}, is extended in this work to account for particles of unequal sizes. This application note outlines the implementation of both the Bagheri approximation \cite{bagheri2024approximate} and the Willett approximation \cite{willett2000capillary} into \textit{MercuryDPM} for modeling liquid capillary bridges. Additionally, we compare the results with the simplified Willett approximation already integrated into \textit{MercuryDPM}. This new implementation significantly improves \textit{MercuryDPM}'s ability to predict the behavior of wet granular materials.

% This application note details the implementation of two liquid capillary bridge fitting formulae, referred to here as the Willett approximation \cite{willett2000capillary} and the Bagheri approximation \cite{bagheri2024approximate}, into \textit{MercuryDPM}. It also compares the results with a simplified Willett approximation already implemented in \textit{MercuryDPM}. While the Bagheri approximation was originally derived for equal-sized spheres, we extend its applicability to particles of unequal sizes in this work. This implementation enhances the predictive power of \textit{MercuryDPM} for simulating wet granular materials.

The following sections of this note describe the particle simulation system and detail the contact models. Subsequently, we demonstrate the implementation of the codes. Finally, we show illustrative examples and discuss the performance and applicability of the approximations in various scenarios. A detailed description of the different capillary force approximations is provided in \ref{A1}.

\section{Software description}
\subsection{Software architecture}
\begin{figure*}[htb!]
\centering
\begin{subfigure}{0.3\linewidth}
    \includegraphics[trim={0cm 0cm 0cm 0cm},clip,width=\linewidth]{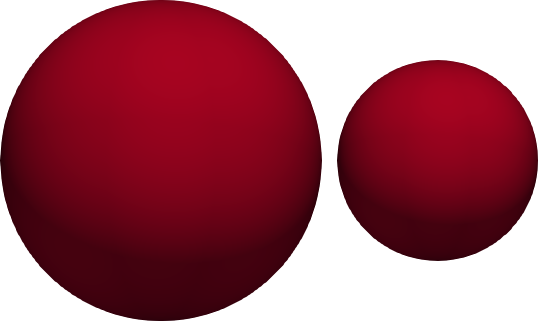}
    \subcaption{}
\end{subfigure}
\hfill
\begin{subfigure}{0.3\linewidth}
    \includegraphics[trim={0cm 0cm 0cm 0cm},clip,width=\linewidth]{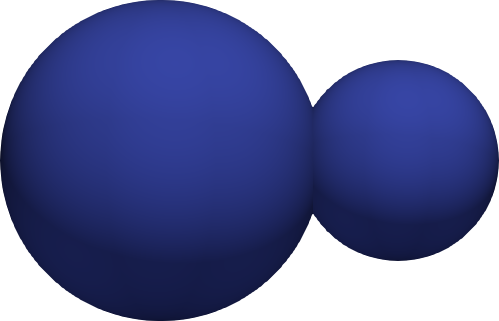}   
    \subcaption{}
\end{subfigure}
\hfill
\begin{subfigure}{0.3\linewidth}
    \includegraphics[trim={0cm 0cm 0cm 0cm},clip,width=\linewidth]{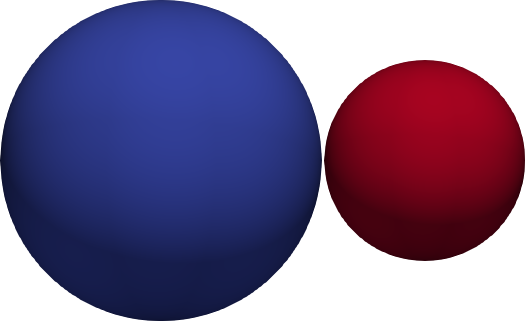}   
    \subcaption{}
\end{subfigure}
\hfill
\begin{subfigure}{0.6\linewidth}
    \includegraphics[trim={0cm 0cm 0cm 0cm},clip,width=\linewidth]{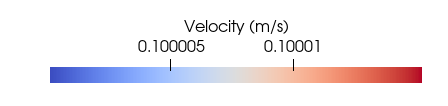}   
\end{subfigure}
\caption{Numerical setup for two-particle collision model for particles of different sizes with bridges (a) before collision (b) during collision and (c) after collision shown for particle sizes $0.50$ mm and $0.80$ mm.}
\label{fig:CollisionModel}
\end{figure*}

Our simulations use a two-particle collision model to analyze interactions between particles of unequal radii, as shown in \autoref{fig:CollisionModel}. The contact model combines dissipative Hertz contact forces \cite{brilliantov1996model,thornton2015granular} with a non-linear liquid bridge model for capillary forces. The results from the implementation of different liquid bridge approximations—namely, the simplified Willett approximation, the classical Willett approximation, and the Bagheri approximation—are compared. The parameters used in DEM simulations are given in \autoref{tab:material_parameters}. Two different particle size ratios are considered in the two-particle simulations: $0.5$ mm and $0.8$ mm, and $0.5$ mm and $1$ mm, respectively. When particles approach each other at a certain distance $S$, they are not linked by a liquid bridge and there is no capillary force. Once particles are in contact, a liquid bridge forms. After contact, the capillary force, $F(S)$, is present at the same distance $S$ between the particles until the bridge ruptures. 

\begin{table}[htbp]
\caption{DEM simulation parameters.}
\label{tab:material_parameters}
\begin{center}
\begin{tabular}{l@{\quad}l@{\quad}ll}
\hline
\multicolumn{1}{l}{\rule{0pt}{18pt}
                   variable}&\multicolumn{1}{l}{unit}&{value}\\[2pt]
                   \hline\rule{0pt}{12pt}\noindent
                   elastic modulus ($E$)  & MPa& 5\\
                   % Restitution coefficient ($\epsilon$)  & -& $0.70$\\
                   % sliding friction coefficient ($\mu$)  & -& 0.30\\
                   Poisson's ratio ($\nu$)  &
                   -& 0.35\\[2pt]
                   particle density ($\rho$)  & kg/m$^3$& 2000\\
                   contact angle ($\theta$)  & degree& $0$\\
                   surface tension ($\gamma$)  & N/m& $0.079$\\
                   liquid bridge volume ($V$)  & nl& $10, 80$\\[2pt] 
\hline                
\end{tabular}
\end{center}
\end{table}
\subsection{Software functionalities}
The simplified Willett approximation is available in \textit{MercuryDPM} under the name \textsf{LiquidBridgeWillet}. The new adhesive contact models are implemented in \textit{MercuryDPM} under the names \textsf{LiquidBridgeBagheri} and \textsf{LiquidBridgeClassicalWillet}, respectively. We compare the force versus interparticle distance results from three liquid bridge approximations for different particle size ratios and different liquid bridge volumes. We discuss the accuracy of each approximation, emphasizing the extended applicability of the Bagheri approximation for simulating polydisperse systems, even though it was originally developed for particles of equal radius.
\subsection{Sample code snippets analysis}
In \ref{A1}, we demonstrate the computation of the capillary forces using the Classical Willett approximation in \textsf{LiquidBridgeClassicalWillet} and the Bagheri approximation in \textsf{LiquidBridgeBagheri}, respectively.

\section{Illustrative examples}
\begin{figure*}[hbt!]
\centering
{\includegraphics[width=0.48\textwidth]{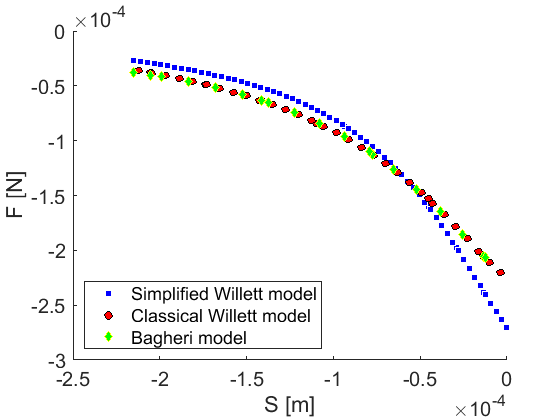}}
\hfill
{\includegraphics[width=0.48\textwidth]{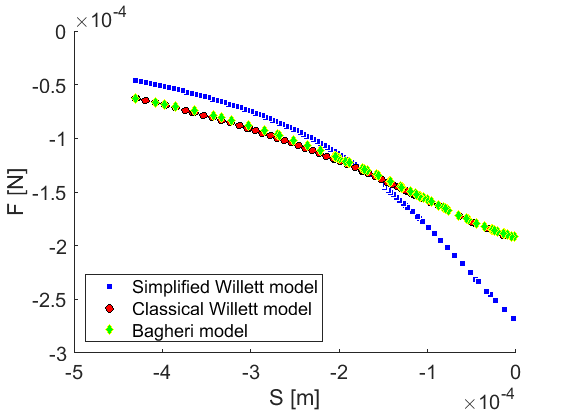}}

\caption{Capillary force, $F$, exerted by a liquid bridge as a function of particle distance $S$ for solutions obtained from the fit equations of simplified Willett approximation, Classical Willett approximation and Bagheri approximation for liquid bridge volume (a) $V = 10$ nl and (b) $V = 80$  nl for particle sizes $0.5$ mm and $0.8$ mm.}

\label{fig:ContactModel}
\end{figure*}

\begin{figure*}[hbt!]
\centering
{\includegraphics[width=0.48\textwidth]{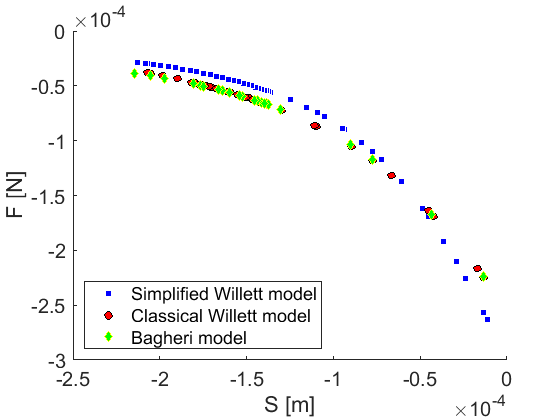}}
\hfill
{\includegraphics[width=0.48\textwidth]{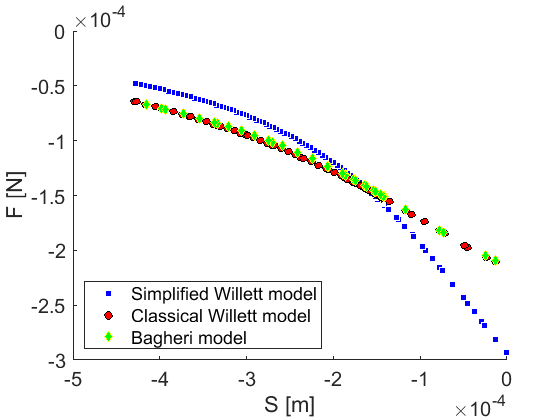}}

\caption{Capillary force, $F$, exerted by a liquid bridge as a function of particle distance $S$ for solutions obtained from the fit equations of simplified Willett approximation, classical Willett approximation and Bagheri approximation for liquid bridge volume (a)$V = 10$ nl and (b) $V = 80$ nl for particle sizes $0.5$ mm and $1$ mm.} 

\label{fig:ContactModel2}
\end{figure*}

As illustrated in \autoref{fig:ContactModel}(a) and (b), the results from the simulation of the two-particle collision model for particle sizes of $0.5$ mm and $0.8$ mm, with liquid bridge volumes of $10$ nl and $80$ nl respectively, reveals an overlap between the forces predicted by the classical Willett approximation and the Bagheri approximation as a function of inter-particle distance. In contrast, the simplified Willett approximation shows a deviation in the capillary force between the two particles for both the liquid bridge volumes. This trend is also observed for particles sized $0.5$ mm and $1$ mm for different liquid bridge volume cases, as shown in \autoref{fig:ContactModel2}(a) and (b). However, the force magnitude increases as the effective interacting particle radius increases in \autoref{fig:ContactModel2}(a) and (b) compared to \autoref{fig:ContactModel}(a) and (b). Thus, the Bagheri approximation is more accurate and closely aligns with the classical Willett approximation, regardless of the liquid bridge volume between particles and particle sizes.

\section{Impact}
All contact models in \textit{MercuryDPM} are defined by a normal, frictional, and adhesive contact models. The normal, frictional and adhesive contact models currently available in \textit{MercuryDPM} are summarised in \cite{weinhart2020fast}. A specific contact model can be chosen by assigning a corresponding species to a particle. The name of the species can be obtained by concatenating the names of the normal, frictional, and adhesive contact model and adding the word \textsf{Species}. For example, users can utilize the interactions provided by particles of type \textsf{HertzianViscoelasticMindlinLiquidBridgeBagheriSpecies}, which include a dissipative Hertz normal force, Mindlin friction, and liquid-bridge adhesion forces based on the Bagheri approximations. Similarly, the adhesive force approximation can be replaced by the classical Willett approximation using interactions of type \textsf{HertzianViscoelasticMindlinLiquidBridgeClassicalWilletSpecies}. The implementations are included in the master branch of \textit{MercuryDPM}. This integration will expand access to these advanced approximations, providing the entire \textit{MercuryDPM} user community with tools to perform more detailed and accurate simulations. 

A detailed description of the test simulation is provided in the \textsf{README.md} file on GitHub. Following these instructions leads to the force calculation shown in \autoref{fig:ContactModel}(a), using the Bagheri model.

\section{Conclusions}

% We have successfully extended the applicability of the Bagheri approximation to particles of unequal sizes and implemented two accurate capillary bridge approximations—the Bagheri approximation and the classical Willett approximation—within the DEM-based open-source code \textit{MercuryDPM}. Through illustrative simulations, we have demonstrated the behavior of these models across various particle size ratios and liquid bridge volumes, and have evaluated their performance and applicability relative to a simplified version of the Willett approximation. We believe that this framework will prove valuable to users of \textit{MercuryDPM}, enabling more precise calculations of capillary forces in particle systems.

We have successfully implemented two accurate capillary bridge approximations—the Bagheri approximation and the classical Willett approximation—within the DEM-based open-source code \textit{MercuryDPM}. The applicability of Bagheri approximation is extended to particles of unequal sizes. Through illustrative simulations, we have demonstrated the behavior of these models across various particle size ratios and liquid bridge volumes, and have evaluated their performance and applicability relative to a simplified version of the Willett approximation. We believe that this framework will prove valuable to users of \textit{MercuryDPM}, enabling more precise calculations of capillary forces in particle systems.

% We have successfully implemented two accurate capillary bridge approximations—the Bagheri approximation and the classical Willett approximation—within the DEM-based open-source code \textit{MercuryDPM}. Through illustrative simulations, we have demonstrated the behavior of these models across various particle size ratios and liquid bridge volumes, and have evaluated their performance and applicability relative to a simplified version of the Willett approximation. We believe that this framework will prove valuable to users of \textit{MercuryDPM}, enabling more precise calculations of capillary forces in particle systems.

% This enhanced framework offers a valuable tool for \textit{MercuryDPM} users, enabling more accurate simulations of capillary interactions in complex particle assemblies, which may have broad implications for research and industrial applications.

\section*{Declaration of competing interest}
There are no conflicts to declare.

\section*{Acknowledgement}
We acknowledge Holger G\"otz for his assistance in integrating the code into the master branch of \textit{MercuryDPM}.

\section*{Data availability}
The implementation of the two models presented here and the test scripts are available at \href{https://github.com/meysam-bagheri/MercuryDPM}{https://github.com/meysam-bagheri/MercuryDPM}.

\appendix
\section{Capillary force approximations}\label{A1}
% \renewcommand{\thesection}{\Alph{section}} % Change the section numbering to letters 
% The parameters of the capillary force approximations $F$, are the contact angle $\theta$, dimensionless liquid volume $V^*=V/R^3$, and the scaled separation distances $S^{+}$ and $S^*$, defined as $S^{+} = S/{{\left(V/R\right)}^{1/2}}$ and $S^* = \left(S/R\right)/{S_c}$, respectively. Here, $V$ is the liquid bridge volume, $S$ is the separation distance and $R$ is the effective particle radius between the interacting particles defined as 

The parameters of the capillary force approximations $F$, are the contact angle $\theta$, liquid bridge volume $V$, and the separation distance $S$. The effective particle radius between the interacting particles $R$, defined as
\begin{equation}
    \frac{2}{R} = \frac{1}{r_1} + \frac{1}{r_2}
    \label{Eq:Harmonic Radius}
\end{equation}
where $r_1$ and $r_2$ are the radius of the two particles $1$ and $2$, respectively. Using the effective radius, we obtain the dimensionless liquid bridge volume $V^*=V/R^3$. The liquid bridge rupture distance, $S_c$ is given by $S_c = \left(1+\theta/2\right)\left({V^*}^{1/3}+\frac{{V^*}^{2/3}}{10}\right)$, and the scaled separation distances $S^{+}$ and $S^*$, defined as $S^{+} = S/{{\left(4V/R\right)}^{1/2}}$  and $S^* = \left(S/R\right)/{S_c}$ , respectively \cite{willett2000capillary,bagheri2024approximate}. The dimensionless capilary force is given by $F^*=F/\left(2\pi R\gamma\right)$. The forces corresponding to all the capillary bridge approximations are active until the liquid bridge ruptures at the distance $S_c$.

% and the scaled separation distances $S^{+}$ and $S^*$, defined as $S^{+} = S/{{\left(V/R\right)}^{1/2}}$ and $S^* = \left(S/R\right)/{S_c}$, respectively.

% The liquid bridge rupture distance, $S_c$ is given by $S_c = \left(1+\theta/2\right)\left({V^*}^{1/3}+\frac{{V^*}^{2/3}}{10}\right)$. The dimensionless capilary force is $F^*=F/R\gamma$. The forces corresponding to all the capillary bridge approximations are active until the liquid bridge ruptures at the distance $S_c$.
\subsection{Simplified Willett approximation}
The simplified Willett approximation \cite{willett2000capillary} which already existed in \textit{MercuryDPM} is defined as follows:
\begin{equation}
    F^* = \frac{\cos\theta}{1+2.1S^{+}+10{S^{+}}^2}
    \label{Eq:Willet_simplified}
\end{equation}

\subsection{Classical Willett approximation}
The classical Willett approximation \cite{bagheri2024approximate} which is now implemented in \textit{MercuryDPM} is defined as follows:
\begin{equation}
    \ln{F^*} = f_1-f_2\exp{\left(f_3\ln{S^{+}}+f_4\ln^2{S^{+}}\right)}
    \label{Eq:Willet_detailed}
\end{equation}
where
\begin{multline*}
 f1 = \left(-0.44507 + 0.050832\theta - 1.1466\theta^2\right) + \\
        \left(-0.1119 - 0.000411\theta - 0.1490\theta^2\right)\ln{V^*} + \\
        \left(-0.012101 - 0.0036456\theta - 0.01255\theta^2\right)\ln^2{V^*} + \\
        \left(-0.0005 - 0.0003505\theta - 0.00029076\theta^2\right)\ln^3{V^*}    
\end{multline*}
\begin{multline*}
f2 = \left(1.9222 - 0.57473\theta - 1.2918\theta^2\right) + \\
        \left(-0.0668 - 0.1201\theta - 0.22574\theta^2\right)\ln{V^*} + \\
        \left(-0.0013375 - 0.0068988\theta - 0.01137\theta^2\right)\ln^2{V^*}  
\end{multline*}
\begin{multline*}
f3 = \left(1.268 - 0.01396\theta - 0.23566\theta^2\right) + \\
        \left(0.198 +0.092\theta - 0.06418\theta^2\right)\ln{V^*} + \\
        \left(0.02232 + 0.02238\theta - 0.009853\theta^2\right)\ln^2{V^*} + \\
        \left(0.0008585 + 0.001318\theta - 0.00053\theta^2\right)\ln^3{V^*}    
\end{multline*}
\begin{multline*}
f4 = \left(-0.010703 + 0.073776\theta - 0.34742\theta^2\right) + \\
        \left(0.03345 + 0.04543\theta - 0.09056\theta^2\right)\ln{V^*} + \\
        \left(0.0018574 + 0.004456\theta - 0.006257\theta^2\right)\ln^2{V^*}    
\end{multline*}

\subsection{Bagheri approximation}
The approximation recently proposed by Bagheri et al. \cite{bagheri2024approximate} which is now implemented in \textit{MercuryDPM} is defined as follows:
\begin{equation}
    {F^*} = F_0^*\frac{1+a_sS^*}{1+ c_{\theta} a_s b_s S^* +c_\theta b_s {S^*}^2}
    \label{Eq:Bagheri}
\end{equation}
where 

\begin{equation*}
    {F_0^*} = \left(1-0.3823{V^*}^{0.2586}\right)\left(1 - a_\theta \sin^{b_\theta}\theta\right)
\end{equation*}

with the following parameters

\begin{equation}
\begin{split}
    {a_{s}} &= -0.3319\,{V^*}^{0.4974}+0.6717\,{V^*}^{0.1995}\\
    {b_{s}} &= 13.84\,{V^*}^{-0.3909}-12.11\,{V^*}^{-0.3945}\,\\
    {c_{\theta}} &= a_c\, \theta^3 + b_c\, \theta + 1\,\\
    {a_{c}} &= -0.007815\,\left(\ln{V^*}\right)^2-0.2105 \,\ln{V^*}-1.426\\
    {b_{c}} &=-1.78\,{V^*}^{0.8351}+0.6669\,{V^*}^{-0.0139}\\
    a_{\theta} &= 0.4158\,V^{*0.2835} +0.6474\\
    b_{\theta} &= -0.2087\,V^{*0.3113} +2.267\,.\\
\end{split}
\label{parameters}
\end{equation}

\section{Code snippets}
\begin{figure}[htb!]
    \centering
    \includegraphics[width=13cm]{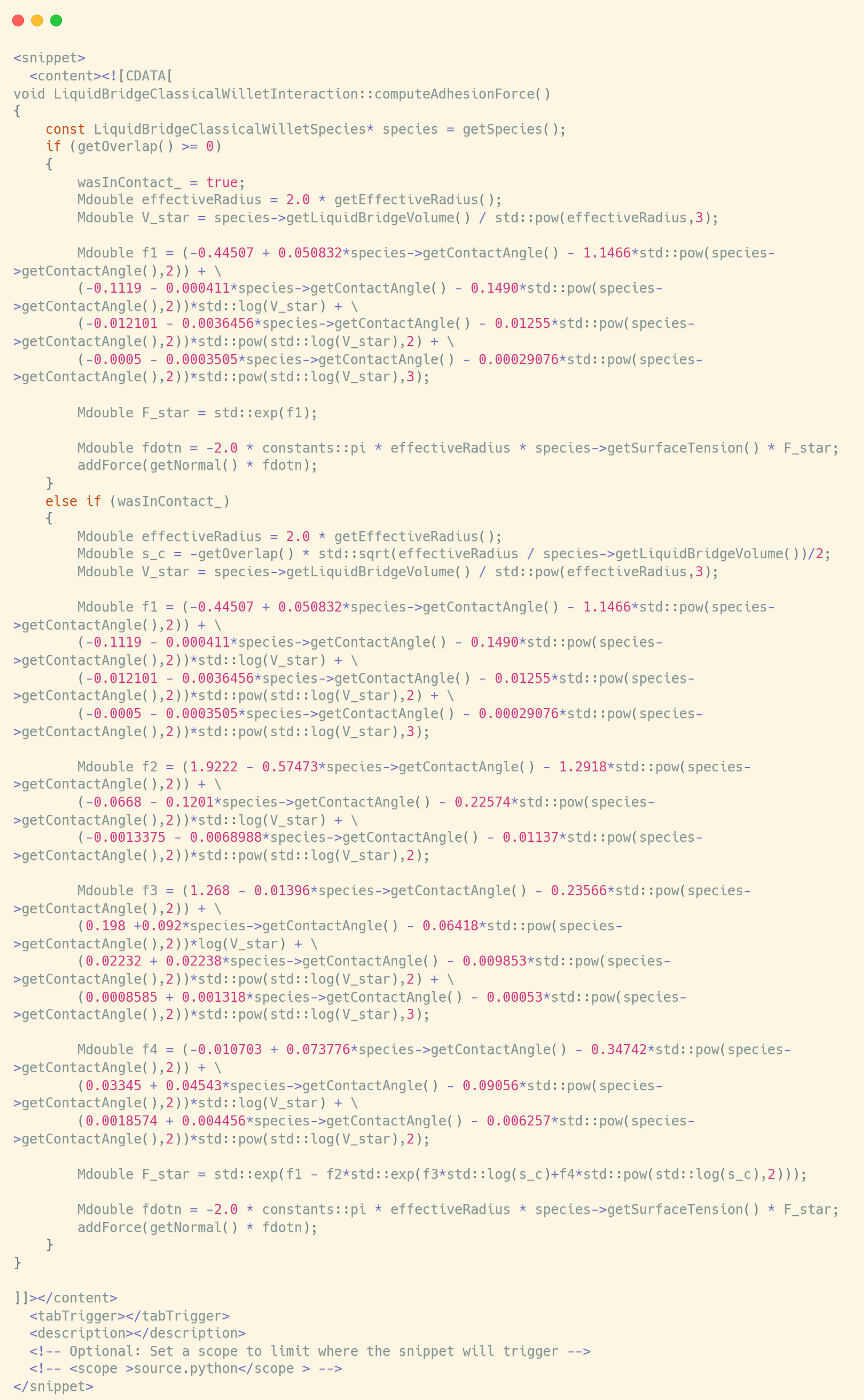}
    \caption{Code snippets for implementation of \textsf{LiquidBridgeClassicalWillet}.}
    \label{fig:SnippetClassicalWillet}
\end{figure}
\begin{figure}[htb!]
    \centering
    \includegraphics[width=13cm]{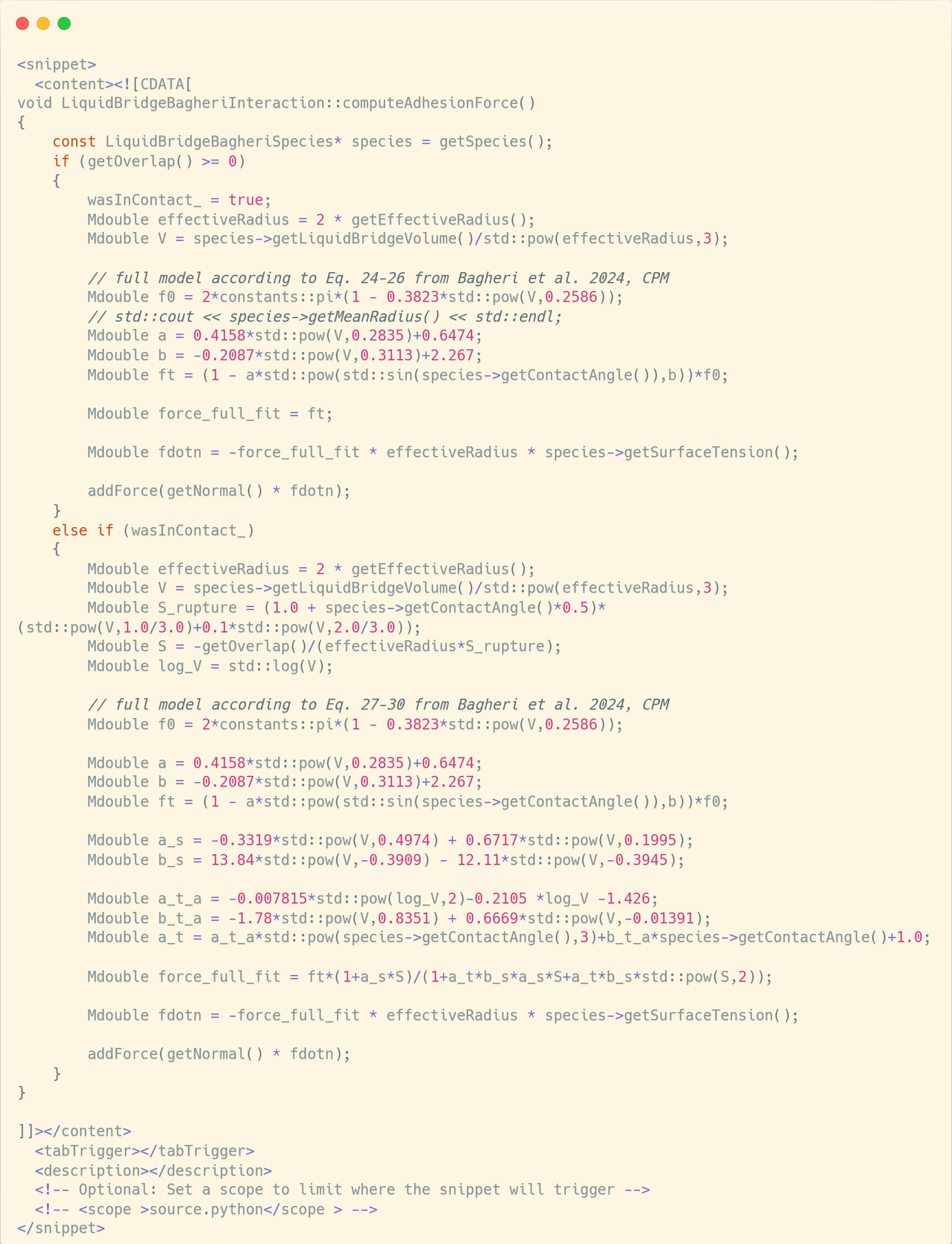}
    \caption{Code snippets for implementation of \textsf{LiquidBridgeBagheri}.}
    \label{fig:SnippetBagheri}
\end{figure}

\bibliography{LiquidBridge}

\end{document}